\documentclass[aps,prc,superscriptaddress,showpacs,nofootinbib,floatfix,twocolumn]{revtex4}
\usepackage{graphicx}

\newcommand{\pt} {\mbox{$p_{T}$}}

\begin{document}

%%%%%%%%%%%%%%%%%%%%%%%%%%%%%%%%%%%%%%%%%%%%%%%%%%%%%%%%%%%%%%%
% Title
%
\title{Nuclear Effects on Hadron Production in 
       d+Au and p+p Collisions at $\sqrt{s_{NN}}$=200 GeV
      }

\newcommand{\abilene}{Abilene Christian University, Abilene, TX 79699, U.S.}
\newcommand{\acadsin}{Institute of Physics, Academia Sinica, Taipei 11529, Taiwan}
\newcommand{\banaras}{Department of Physics, Banaras Hindu University, Varanasi 221005, India}
\newcommand{\barc}{Bhabha Atomic Research Centre, Bombay 400 085, India}
\newcommand{\bnl}{Brookhaven National Laboratory, Upton, NY 11973-5000, U.S.}
\newcommand{\caucr}{University of California - Riverside, Riverside, CA 92521, U.S.}
\newcommand{\ciae}{China Institute of Atomic Energy (CIAE), Beijing, People's Republic of China}
\newcommand{\cns}{Center for Nuclear Study, Graduate School of Science, University of Tokyo, 7-3-1 Hongo, Bunkyo, Tokyo 113-0033, Japan}
\newcommand{\colorado}{University of Colorado, Boulder, CO 80309, U.S.}
\newcommand{\columbia}{Columbia University, New York, NY 10027 and Nevis Laboratories, Irvington, NY 10533, U.S.}
\newcommand{\dapnia}{Dapnia, CEA Saclay, F-91191, Gif-sur-Yvette, France}
\newcommand{\debrecen}{Debrecen University, H-4010 Debrecen, Egyetem t{\'e}r 1, Hungary}
\newcommand{\elte}{ELTE, E{\"o}tv{\"o}s Lor{\'a}nd University, H - 1117 Budapest, P{\'a}zm{\'a}ny P. s. 1/A, Hungary}
\newcommand{\fsu}{Florida State University, Tallahassee, FL 32306, U.S.}
\newcommand{\gsu}{Georgia State University, Atlanta, GA 30303, U.S.}
\newcommand{\hiroshima}{Hiroshima University, Kagamiyama, Higashi-Hiroshima 739-8526, Japan}
\newcommand{\ihepprot}{IHEP Protvino, State Research Center of Russian Federation, Institute for High Energy Physics, Protvino, 142281, Russia}
\newcommand{\illuiuc}{University of Illinois at Urbana-Champaign, Urbana, IL 61801, U.S.}
\newcommand{\isu}{Iowa State University, Ames, IA 50011, U.S.}
\newcommand{\jinrdubna}{Joint Institute for Nuclear Research, 141980 Dubna, Moscow Region, Russia}
\newcommand{\kek}{KEK, High Energy Accelerator Research Organization, Tsukuba, Ibaraki 305-0801, Japan}
\newcommand{\kfki}{KFKI Research Institute for Particle and Nuclear Physics of the Hungarian Academy of Sciences (MTA KFKI RMKI), H-1525 Budapest 114, POBox 49, Budapest, Hungary}
\newcommand{\korea}{Korea University, Seoul, 136-701, Korea}
\newcommand{\kurchatov}{Russian Research Center ``Kurchatov Institute", Moscow, Russia}
\newcommand{\kyoto}{Kyoto University, Kyoto 606-8502, Japan}
\newcommand{\labllr}{Laboratoire Leprince-Ringuet, Ecole Polytechnique, CNRS-IN2P3, Route de Saclay, F-91128, Palaiseau, France}
\newcommand{\lawllnl}{Lawrence Livermore National Laboratory, Livermore, CA 94550, U.S.}
\newcommand{\losalamos}{Los Alamos National Laboratory, Los Alamos, NM 87545, U.S.}
\newcommand{\lpc}{LPC, Universit{\'e} Blaise Pascal, CNRS-IN2P3, Clermont-Fd, 63177 Aubiere Cedex, France}
\newcommand{\lund}{Department of Physics, Lund University, Box 118, SE-221 00 Lund, Sweden}
\newcommand{\muenster}{Institut f\"ur Kernphysik, University of Muenster, D-48149 Muenster, Germany}
\newcommand{\myongji}{Myongji University, Yongin, Kyonggido 449-728, Korea}
\newcommand{\nagasaki}{Nagasaki Institute of Applied Science, Nagasaki-shi, Nagasaki 851-0193, Japan}
\newcommand{\newmex}{University of New Mexico, Albuquerque, NM 87131, U.S. }
\newcommand{\nmsu}{New Mexico State University, Las Cruces, NM 88003, U.S.}
\newcommand{\ornl}{Oak Ridge National Laboratory, Oak Ridge, TN 37831, U.S.}
\newcommand{\orsay}{IPN-Orsay, Universite Paris Sud, CNRS-IN2P3, BP1, F-91406, Orsay, France}
\newcommand{\peking}{Peking University, Beijing, People's Republic of China}
\newcommand{\pnpi}{PNPI, Petersburg Nuclear Physics Institute, Gatchina, Leningrad region, 188300, Russia}
\newcommand{\riken}{RIKEN (The Institute of Physical and Chemical Research), Wako, Saitama 351-0198, JAPAN}
\newcommand{\rikjrbrc}{RIKEN BNL Research Center, Brookhaven National Laboratory, Upton, NY 11973-5000, U.S.}
\newcommand{\saopaulo}{Universidade de S{\~a}o Paulo, Instituto de F\'{\i}sica, Caixa Postal 66318, S{\~a}o Paulo CEP05315-970, Brazil}
\newcommand{\seoulnat}{System Electronics Laboratory, Seoul National University, Seoul, South Korea}
\newcommand{\stonybrkc}{Chemistry Department, Stony Brook University, SUNY, Stony Brook, NY 11794-3400, U.S.}
\newcommand{\stonycrkp}{Department of Physics and Astronomy, Stony Brook University, SUNY, Stony Brook, NY 11794, U.S.}
\newcommand{\subatech}{SUBATECH (Ecole des Mines de Nantes, CNRS-IN2P3, Universit{\'e} de Nantes) BP 20722 - 44307, Nantes, France}
\newcommand{\tenn}{University of Tennessee, Knoxville, TN 37996, U.S.}
\newcommand{\titech}{Department of Physics, Tokyo Institute of Technology, Oh-okayama, Meguro, Tokyo 152-8551, Japan}
\newcommand{\tsukuba}{Institute of Physics, University of Tsukuba, Tsukuba, Ibaraki 305, Japan}
\newcommand{\vandy}{Vanderbilt University, Nashville, TN 37235, U.S.}
\newcommand{\waseda}{Waseda University, Advanced Research Institute for Science and Engineering, 17 Kikui-cho, Shinjuku-ku, Tokyo 162-0044, Japan}
\newcommand{\weizmann}{Weizmann Institute, Rehovot 76100, Israel}
\newcommand{\yonsei}{Yonsei University, IPAP, Seoul 120-749, Korea}
\newcommand{\deceased}{\dagger}
\affiliation{\abilene}
\affiliation{\acadsin}
\affiliation{\banaras}
\affiliation{\barc}
\affiliation{\bnl}
\affiliation{\caucr}
\affiliation{\ciae}
\affiliation{\cns}
\affiliation{\colorado}
\affiliation{\columbia}
\affiliation{\dapnia}
\affiliation{\debrecen}
\affiliation{\elte}
\affiliation{\fsu}
\affiliation{\gsu}
\affiliation{\hiroshima}
\affiliation{\ihepprot}
\affiliation{\illuiuc}
\affiliation{\isu}
\affiliation{\jinrdubna}
\affiliation{\kek}
\affiliation{\kfki}
\affiliation{\korea}
\affiliation{\kurchatov}
\affiliation{\kyoto}
\affiliation{\labllr}
\affiliation{\lawllnl}
\affiliation{\losalamos}
\affiliation{\lpc}
\affiliation{\lund}
\affiliation{\muenster}
\affiliation{\myongji}
\affiliation{\nagasaki}
\affiliation{\newmex}
\affiliation{\nmsu}
\affiliation{\ornl}
\affiliation{\orsay}
\affiliation{\peking}
\affiliation{\pnpi}
\affiliation{\riken}
\affiliation{\rikjrbrc}
\affiliation{\saopaulo}
\affiliation{\seoulnat}
\affiliation{\stonybrkc}
\affiliation{\stonycrkp}
\affiliation{\subatech}
\affiliation{\tenn}
\affiliation{\titech}
\affiliation{\tsukuba}
\affiliation{\vandy}
\affiliation{\waseda}
\affiliation{\weizmann}
\affiliation{\yonsei}
\author{S.S.~Adler}	\affiliation{\bnl}
\author{S.~Afanasiev}	\affiliation{\jinrdubna}
\author{C.~Aidala}	\affiliation{\columbia}
\author{N.N.~Ajitanand}	\affiliation{\stonybrkc}
\author{Y.~Akiba}	\affiliation{\kek} \affiliation{\riken}
\author{A.~Al-Jamel}	\affiliation{\nmsu}
\author{J.~Alexander}	\affiliation{\stonybrkc}
\author{K.~Aoki}	\affiliation{\kyoto}
\author{L.~Aphecetche}	\affiliation{\subatech}
\author{R.~Armendariz}	\affiliation{\nmsu}
% ...
\collaboration{PHENIX Collaboration} \noaffiliation

\date{\today}

%%%%%%%%%%%%%%%%%%%%%%%%%%%%%%%%%%%%%%%%%%%%%%%%%%%%%%%%%%%%%%%
% Abstract
%

\begin{abstract}

%Enhanced hadron production in p+A collisions relative to p+p collisions is generally known
%as the Cronin effect, and is studied by 

PHENIX has measured the centrality dependence of mid-rapidity
pion, kaon and proton transverse momentum distributions in d+Au 
and p+p collisions at $\sqrt{s_{NN}}$ =  200 GeV.  
The p+p data provide a reference for 
nuclear effects in d+Au and previously measured Au+Au collisions.
Hadron production is enhanced in d+Au, relative to independent
nucleon-nucleon scattering, as was observed
in lower energy collisions.
The nuclear modification factor for (anti) protons is larger
than that for pions. The difference increases with centrality, but 
is not sufficient to account for
the  abundance of baryon production observed in central Au+Au
collisions at RHIC.
The centrality dependence in d+Au shows that the nuclear 
modification factor increases gradually with the number of 
collisions suffered by each participant nucleon.
We also present comparisons with lower energy 
data as well as with parton recombination and other theoretical
models of nuclear effects on particle production.
%The nuclear modification factors and particle ratios
%indicate significant enhancement of baryon production in d+Au collisions, 
%albeit insufficient to explain the ''anomalous" large
%proton to pion ratio in central Au+Au collisions. Pions show a smaller enhancement.

\end{abstract}

\maketitle

%%%%%%%%%%%\begin{multicols}{2}   % This is needed only for the "multicols" style
%%%%%%%%%%%\narrowtext            % This is needed only for the "multicols" style
%%%%%%%%%%%%%%%%%%%%%%%%%%%%%%%%%%%%%%%%%%%%%%%%%%%%%%%%%%%%%
%NOTES: 
%
%1.  For PRL do not use section headings.
%
%2.  Do not worry about indenting the first line of a paragraph.  Just
%    insert a blank line between paragraphs.  Similarly, if you want 
%    an equation to stay within a paragraph, do not put a blank line
%    before or after the equation.
%
%3.  Do not imbed figures or tables; place them all at the end (see below).
%
%4.  Name all references and use "\cite{refname}" in the text to cite them.
%    (The RevTeX macro will replace this with "[1]" in proper PR style.)
%
%5.  The list of references must be ordered in the same sequence as they
%    occur in the text.
%
%6.  Use our standard aknowldegement below as the last paragraph of your
%    text.  (Yes, it does count toward the length!).
%
%%%%%%%%%%%%%%%%%%%%%%%%%%%%%%%%%%%%%%%%%%%%%%%%%%%%%%%%%%%%%
% general introduction
%
% \marginpar{{\small \em Intro}}
%
%\twocolumn

\section {Introduction}

Since the early 1970's, it is well established 
\cite{cronin,straub,antreasyan}, 
that energetic particle production in proton-nucleus (p+A) collisions 
%does not scale with 
increases faster than the number of 
%available 
binary nucleon-nucleon collisions. This effect, called the ``Cronin effect'', 
is a manifestation of the fact that particle production 
and propagation is influenced by 
%the color field inside 
the nucleus.
%More specifically, i
If the A-dependence of the 
invariant cross section, $I$,  of particle $i$ in p+A collisions
is parameterized as
\begin{equation}
I_{i}(\pt,A) = I_{i}(\pt,1)\cdot A^{\alpha_{i}(\pt)}
\label{eqn:Cronin}
\end{equation}
%where the index $i$ refers to the particle species produced,
then it has been observed that $\alpha_{i}$ is greater than unity
above some transverse momentum value, typically \mbox{1-1.5 GeV/c}, 
denoting significant enhancement of particle production in p+A collisions. 
The enhancement depends on the momentum and the type of particle produced, 
with protons and antiprotons exhibiting a much larger 
enhancement than pions and kaons at $p_T > 2-3$ GeV/c. At $\sqrt{s_{NN}}$ =  27.4 GeV, 
the enhancement peaks at around \pt $=$4.5 GeV/c, 
with \mbox{$\alpha_{K^{+}} \simeq  \alpha_{\pi^{+}} = 1.109 \pm 0.007$}, while, at the
same momentum, the protons can be described by an $\alpha$-factor of 
$\alpha_{p}-\alpha_{\pi^{+}}=0.231 \pm 0.013$ \cite{antreasyan}.   

Although the observables in Eq.~\ref{eqn:Cronin} have been clearly related to 
the nuclear medium, the cause of the Cronin enhancement and its species dependence are not yet 
completely understood and further experimental study is 
%interesting 
warranted in its own right. Furthermore, 
in the search for the Quark Gluon Plasma at the Relativistic Heavy Ion Collider (RHIC), 
the Cronin effect is extremely important, as
%acquires new levels of significance. 
novel effects observed in central Au+Au collisions
require good control 
of the initial state conditions. At RHIC energies, it was discovered 
that hadron production at high transverse momentum ($p_T \ge 2$\,GeV/$c$) is suppressed in central 
Au+Au collisions \cite{ppg003} compared to nucleon-nucleon collisions. 
Such suppression may be interpreted as a consequence of the energy loss suffered by the 
hard-scattered partons as they propagate through the hot and dense medium.
However, since the Cronin effect acts in the opposite direction enhancing the hadron 
yields, it has to be taken into account when the parton energy loss is  
determined from the data.

Another discovery at RHIC, originally unexpected, is that the yields of $p$ and 
$\bar{p}$ at intermediate \pt\ ($ 1.5 < \pt < 5$\,GeV/$c$ ) in 
central Au+Au collisions \cite{ppg006,ppg015,ppg026} are comparable to the yield of pions, in striking contrast 
to the proton to pion ($p/\pi$) ratios of $\sim$ 0.1 - 0.3 measured in p+p 
collisions \cite{ISR}. Novel mechanisms of particle production in the environment of 
dense matter, such as recombination of boosted quarks \cite{recomb}
or contributions from baryon junctions \cite{junct}, which can become dominant 
in the presence of pion suppression were proposed to explain the data. Since it has 
been observed that at lower energies Cronin enhancement is stronger for 
protons than for pions \cite{antreasyan}, this effect has to be considered at RHIC 
before new physics is invoked.

The effects from the initial state are best studied by performing 
a control experiment in which 
%only cold nuclear matter is produced. 
no hot and dense matter is produced. Deuteron + gold collisions 
at $\sqrt{s_{NN}}$= 200 GeV serve this purpose. Since 
there is no  hot and dense final state 
medium,
%is removed, 
the initial state conditions become accessible 
to the experiment. In addition to Cronin enhancement, known initial 
state effects also include nuclear shadowing 
and gluon saturation \cite{colorglas}.   
The Cronin enhancement is usually attributed to momentum broadening due to multiple
initial state soft \cite{LevPetersson} or semi-hard \cite{Accardi, PappLevai, Vitev, Wang} scattering. 
Such models typically do not predict the particle species 
dependence observed in the data.  
Recently, Hwa and collaborators provided an alternative explanation due to final state interactions. 
The particle species dependent enhancement is attributed to recombination 
of shower quarks with those from the medium, where no distinction is made if hot or 
cold nuclear  matter is produced \cite{Hwa}. 
%Currently, the theoretical  explanation 
%of the Cronin enhancement is at best incomplete. 
Identified hadron production 
measured as a function of centrality brings important experimental 
%handles 
data relevant to the long outstanding problem of the baryon Cronin effect. 
The dependence of the enhancement upon 
the thickness of the medium, or the number of collisions suffered by each 
participating nucleon, can help
differentiate among the different scattering models, and 
the species dependence helps to separate initial from final state effects in d+Au.

The paper is arranged as follows. Section II describes the experiment, 
data analysis, and systematic uncertainties. 
Section III presents hadron spectra, yields
and the resulting nuclear modification factors. Discussion of the
centrality, energy and species dependence of the nuclear modification
factors and implications for understanding of the Cronin effect
are in section IV. Section V presents conclusions.

%%%%%%%%%%%%%%%%%%%%%%%%%%%%%%%%%%%%%%%%%%%%%%%%%%%%%%%%%%%%%%%%%%%%%%%%
\section {Experiment and Data Analysis}
%%%%%%%%%%%%%%%%%%%%%%%%%%%%%%%%%%%%%%%%%%%%%%%%%%%%%%%%%%%%%%%%%%%%%%%%

%%%%%%%%%%%%%%%%%%%%%%%%%%%%%%%%%%%%%%%%%%%Table I
 \begin{table*}[tbh]
  \caption{Mean number of binary collisions, participating nucleons from the Au nucleus,
     number of collisions per participating deuteron nucleon,
    and trigger bias corrections for the d+Au centrality bins.}
  \label{table:ncoll}
 % \begin{center}
\begin{ruledtabular}
    \begin{tabular}{ccccc}
                   & 00-20$\%$ &  20-40$\%$ & 40-60$\%$ &  60-88$\%$ \\ \hline
         $\langle N_{coll} \rangle$   &  15.4$\pm$1.0    &    10.6$\pm$0.7   &  7.0$\pm$0.6      &   3.1$\pm$0.3  \\
         $\langle N_{part} \rangle$   &  15.6$\pm$0.9    &    11.1$\pm$0.6   &  7.7$\pm$0.4      &   4.2$\pm$0.3  \\
         $\langle \nu = N_{coll}/N_{part}^d \rangle$ &  7.5$\pm$0.5    &    5.6$\pm$0.4   &  4.0$\pm$0.3     &   
2.2$\pm$0.2  \\
Trigger bias correction & 0.95$\pm$0.3 & 0.99$\pm0.007$ & 1.03$\pm$0.009 & 1.04$\pm$0.027\\
      \end{tabular}
\end{ruledtabular}
%    \end{center}
  \end{table*}

\subsection {Data Sets and Trigger}
\label{subsec:exp}

Data presented here include collisions at $\sqrt{s_{NN}} = 200$~GeV 
of Au+Au taken in the 2002 run of RHIC and d+Au and p+p collected
in 2003. In the following we discuss analysis of the p+p and d+Au data;
details of the Au+Au analysis, and the Au+Au results, are found 
in \cite{ppg026}. 
Events with vertex position along the beam axis within  $|z| < $30 cm
were triggered by the Beam-Beam  Counters (BBC) located 
at $|\eta|$~=~3.0-3.9 \cite{nim_phenix}. The minimum bias trigger 
accepts 88.5 $\pm$ 4\% of all d+Au collisions that satisfy the vertex condition, 
and 51.6 $\pm$ 9.8\% of p+p collisions. A total of 42 $\times$ 10$^6$ 
minimum bias d+Au events 
and 25 $\times$ 10$^6$ minimum bias p+p events were analyzed.

In p+p collisions, PHENIX determines the differential invariant cross 
section via
%without assumptions for any of the terms in 
\begin{equation}
E\frac{ d^{3}\sigma } { dp^{3} } =  \frac{\sigma_{BBC}}{N^{Total}_{BBC}}  \cdot \frac{1}{2\pi} \cdot \frac{1}{p_T} 
                           \cdot C^{geo}_{eff}(\pt)  \cdot C^{BBC}_{bias} \cdot  \frac{d^2N}{ dp_{T} dy } \, \,  \, \,.
\label{eq:goldenformula}
\end{equation}

The BBC cross section $\sigma_{BBC}$ was 
%calculated by counting the BBC rates during a vernier scan, 
determined via the van der Meer scan technique \cite{vanderMeer}. 
%when the colliding beams have maximum overlap; 
In this p+p data set,
%and, as mentioned previously, it was measured to be 
$\sigma_{BBC} =  23.0 \, \pm \, 2.2(9.6\%) $ mb.
$N^{Total}_{BBC}$ is the total number of BBC triggers analyzed.
The factor $C^{geo}_{eff}(\pt)$ 
denotes the efficiency and geometrical acceptance correction, 
calculated with a detailed GEANT Monte Carlo simulation \cite{GEANT}
of the PHENIX detector.
$C^{geo}_{eff}(\pt)$ 
normalizes the cross section in one unit of rapidity and full azimuthal coverage.
The $C^{BBC}_{bias}$ factor
corrects for the fact that the forward BBC trigger counters
measure only a fraction of the inelastic p+p cross section. This subset of 
events on which the BBC triggers contains only a fraction of the inclusive
particle yield at mid-rapidity. 
%This fraction is determined from data collected
%by a separate trigger, independent of the BBC, and is described
%in detail in \cite{ppg024} for neutral pions using the Electromagnetic Calorimeters.
For charged hadrons this fraction was determined using triggers 
on the beam crossing clock; the fraction was found
to be 
%the same with the fraction reported in \cite{ppg024}, 
0.80 $\pm$ 0.02, independent of \pt. The
$C^{BBC}_{bias}$ term is, in our nomenclature, the inverse of this fraction.

%There were no van der Meer scans performed during the 
%d+Au run, therefore PHENIX
%only measures the inelastic yield per BBC triggered event. 
In d+Au collisions PHENIX measures the inelastic 
yield per BBC triggered event.\footnote{To convert the reported
inelastic yield to differential cross section, one must multiply
by the beam-beam trigger cross section of 
$\sigma^{tot}_{MB}(d\rm{Au})\epsilon^{BBC}_{MB}(d\rm{Au}) = 
\rm{1.99 \pm 0.10 b}$ \cite{jpsi_paper}.}
The collision centrality is selected in d+Au using the south 
(Au-going side) BBC (BBCS). We assume that the BBCS signal is 
proportional to the number of participating nucleons ($N^{Au}_{part}$) 
in the Au nucleus, and that the hits in the BBCS are uncorrelated 
to each other. 
%%GLAUBER MODEL DETAILS START HERE ------------------------
We use a Glauber model \cite{glauber} and simulation 
of the BBC to define 4 centrality classes in d+Au collisions, as 
discussed in detail in \cite{ppg036}. The deuteron nucleus is modeled 
after a wave function due to Hulth\`en \cite{hulthen}

\begin{equation}
  \label{eq:hulthen}
  \phi_\mathrm d(\mathbf r_\mathrm{pn}) =
  \left(\frac{\alpha\beta(\alpha+\beta)}{2\pi(\alpha-\beta)^2}\right)^{\frac{1}{2}}(\mathrm e^{-\alpha r_\mathrm{pn}} - \mathrm e^{-\beta r_\mathrm{pn}}) / r_\mathrm{pn}
\end{equation} 
where $\alpha = 0.228 \, \mathrm {fm}^{-1}$ and $\beta = 1.18 \,
\mathrm {fm}^{-1}$ and $\mathbf r_{\mathrm{np}}$
refers to the separation between the proton and the neutron.
The Au nucleus is modeled using a Woods-Saxon density distribution
\begin{equation}
  \label{eq:woods-saxon}
  \rho(r) = \frac{\rho_0}{1+\exp\left(\frac{r-c}{a}\right)}
\end{equation}
where $a = 0.54$\,fm and $c = 1.12 \times A^{1/3} - 0.86 \times
A^{-1/3} = 6.40$\,fm, a value which agrees well with the measured
charge radius of $R = 6.38$\,fm for gold.
%%GLAUBER MODEL DETAILS FINISH HERE ------------------------

Using the above parameters and taking into account the BBC efficiency the mean number of binary
collisions along with the mean number of participating nucleons
from the Au nucleus that correspond to each centrality bin are shown in Table I.
For the minimum bias d+Au collisions, $\langle N_{coll} \rangle$ = 8.5 $\pm$ 0.4
and  $\langle N_{part} \rangle$ = 9.1 $\pm$ 0.4.

As for p+p collisions, there is a BBC trigger bias,
but it is much smaller in d+Au. 
In addition, a second bias occurs in d+Au centrality selected collisions. 
%which we call it the bin-shifting effect,
This second bias arises from the fact that events containing
high \pt\ hadrons from hard scatterings may have larger multiplicity,
and consequently produce a larger signal in the BBCS.
Such events would be considered more central than 
events without a hard scattering. This effect gives an 
opposite bias from the first trigger bias effect in the most
peripheral bin as events can be shifted out of this bin but not
into it.
%for more peripheral centrality bins. 
We correct for both biases in 
d+Au collisions using simulations and a Glauber model. The combined 
corrections for these effects range from 0-5\%, depending on the 
centrality category, and are shown in Table I. 
Systematic uncertainties on these corrections are less than 4\%.

%%%%%%%%%%%%%%%%%%%%%%%%%%%%%%%%%%%%%%

\subsection {Tracking and Particle Identification}

Charged particles are reconstructed using a drift chamber (DC) and 
two layers of multi-wire proportional chambers with pad readout
(PC1, PC3) \cite{nim_phenix}. Pattern recognition is based on a 
combinatorial Hough transform in the track bend plane, while the 
polar angle is determined by PC1 and the location 
of the collision vertex along the beam direction~\cite{trackingNIM}. 
The track reconstruction efficiency is approximately 98\%.

Particle momenta are measured with a resolution 
$\delta p/p = 0.7\% \oplus 1.1\%p~(GeV/c)$. 
The momentum scale is known to 0.7\%. Particle identification is 
based on particle mass calculated from the measured momentum and the 
velocity obtained from the time-of-flight and path length along the 
trajectory. The measurement uses the 
portion of the spectrometer containing the high resolution
time-of-flight(TOF) detector, which covers pseudo-rapidity 
-0.35 $\le \eta \le 0.35$ and $\Delta\phi=\pi/8$ in azimuthal angle.    
The timing uses the BBC for the global start, and stop signals
from the TOF scintillators located at a radial distance of 5.06 m.
%% addition form BVJ to address confusion about track matchin
%%  expressed by Akiba
%%
Tracks are required to match a hit on the TOF within $\pm$ 
3 standard deviations ($\sigma$) of the track projection to 
the TOF radial location.
The particle yields are corrected for losses due to this matching
cut.

\begin{figure}
\begin{center}
\includegraphics[width=1.0\linewidth]{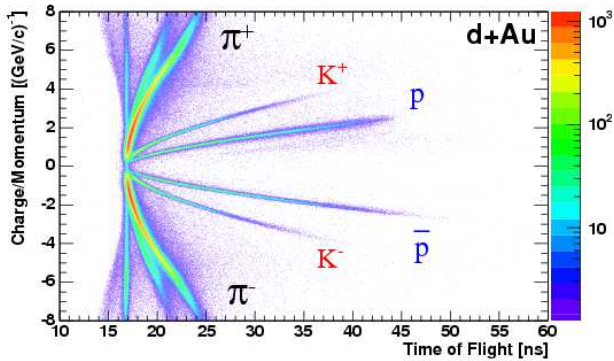}
\caption{(color online) Time of flight vs. charge/momentum for particles in 
minimum bias d+Au collisions. Bands corresponding to the
different hadrons are labeled. Electron and muon bands
are also visible, though not labeled.
}
\label{fig:pid} 
\end{center}
\end{figure}

\begin{figure}
\begin{center}
\includegraphics[width=1.0\linewidth]{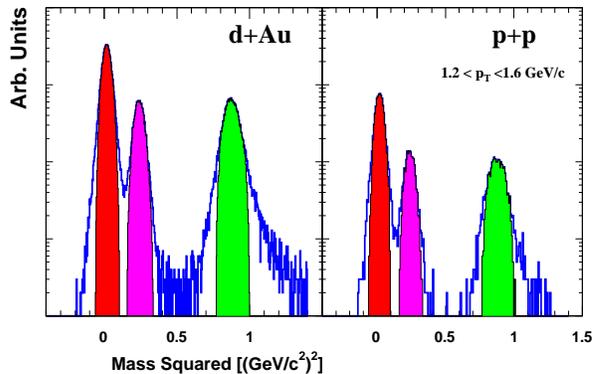}

\caption{(color online) Mass squared distribution for positively charged tracks
with 1.2 $\le p_T \le$ 1.6 GeV/$c$, using the high 
resolution time-of-flight measurement. The solid regions indicate the 
mass squared ranges for accepted pions, kaons and 
protons, respectively, from left to right.
}
\label{fig:mass2} 
\end{center}
\end{figure}

Figure \ref{fig:pid} illustrates the performance of the hadron
identification system.
The system resolution is $\sigma \approx$ 130 ps for both d+Au and
p+p; 
the start time resolution is limited by the low hit multiplicity in
the BBC. The TOF resolution achieved allows for a
clear separation of $\pi/K$ and 
$K/p$ up to $p_T$ = 2 GeV/c and $p_T$ = 3.6 GeV/c, respectively. 
A $2 \sigma$ cut in mass squared is used to separate the different 
hadron species, as shown in Fig.~\ref{fig:mass2} for hadrons in
the $p_T$ range from 1.2 - 1.6 GeV/c. A clear separation between
pions and kaons is seen. The particle yields are corrected for
losses due to the $2 \sigma$ cut.
%%%%%%%%%%%%%%%%%%%%%%%%%%%%%%%%%%%%%%

%%%%%%%%%%%%%%%%%%%%%%%%%%%%%%%%%%%%%%%%  fig:pid

%%%%%%%%%%%%%%%%%%%%%%%%%%%%%%%%%%%%%%%% fig:mass2

Corrections to the charged particle spectrum for geometrical 
acceptance, decays in flight, reconstruction
efficiency, energy loss in detector material,
and momentum resolution are determined using a 
single-particle GEANT Monte Carlo simulation. 

\begin{figure}
\begin{center}
\includegraphics[width=1.0\linewidth]{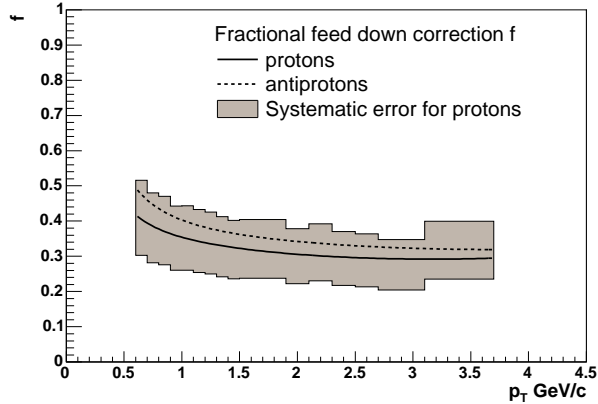}

\caption{Fractional contributions of protons, f, (solid line) and 
antiprotons (dashed line), as a function of \pt, from weak decays in all measured
protons and antiprotons. Systematic error band (26$\%$) for 
the protons is shown and discussed in the text. The same systematic error 
applies for the antiprotons.
}
\label{fig:feedplot} 
\end{center}
\end{figure}

The proton and antiproton spectra are corrected for feed-down
from weak decays via a Monte Carlo simulation using as input
experimental data on $\Lambda$ production. 
 The total number of protons produced in the collisions
can be written as: 
 $p+0.64(\Lambda + \Sigma^0 + \Xi^0 + \Xi^- + \Omega^-) + 0.52\Sigma^+$ 
where $p$ denotes the primordial number of protons produced, and the other symbols
denote the primordial number of those particles produced in the collision.
$0.64$ and $0.52$ are the branching ratios for
$\Lambda\rightarrow p\pi^-$ and $\Sigma^+\rightarrow p\pi^0$, respectively.
The hyperons listed together with $\Lambda$ decay to $\Lambda$ with 
approximately 100\% branching ratio, and so yield protons with the 
$\Lambda\rightarrow p\pi$ branching ratio.
 We estimate the proton and antiproton spectra from weak decays and subtract 
them from the measured yields, using
experimental data from UA5 \cite{UA5} 
from non-single diffractive $\sqrt{s}$ = 200 GeV 
 $p+\bar{p}$ collisions, and preliminary $\Lambda$ and $\overline\Lambda $
spectra from STAR 
in p+p and d+Au collisions at the same 
energy \cite{heinzpaper, heinzppt, cai, mtscaling}.
 The shape of the $\Sigma^0, \Xi^0$, and $\Xi^-$
 spectra are constructed from the $\Lambda$ spectrum by $m_T$ scaling
(i.e. under the assumption that these hadrons are all produced with
roughly the same spectrum in transverse mass \cite{mtscaling}). 
%This approach is justified by
% \cite{mtscaling}, where the mt-scaled spectra 
%of these particles collapse to a universal shape,
% although the absolute normalization of mt-scaling at these energies 
%breaks down as was shown at \cite{mtscaling}. 
The relative normalization of $\Lambda, \Sigma^0, \Xi^0$, and $\Xi^-$
from UA5 \cite{UA5} is used for both p+p and d+Au collisions. 
 This is justified by the similarity of the Cronin effect for different 
baryons. The contribution of $\Omega^-$ is negligible, and is 
not included in the correction.
The Monte Carlo simulation decays these baryons and propagates the products 
through the PHENIX magnetic field and central arm detectors, accounting also for
the change of momentum distributions between parent and daughter particles 
due to decay kinematics. 
The resulting proton and antiproton spectra are then subtracted from the 
measured inclusive spectra. 
 The fractional contribution from feed-down protons from weak decays
to the total measured proton spectrum, $f$, is approximately 
30$\%$ at moderate and high \pt\, growing slowly for lower $p_T$
and reaching 40$\%$ at $p_T$ = 0.6 GeV/c. $f$ is shown as a function
of \pt\ in Fig.~\ref{fig:feedplot}, along with the corresponding
fractional contribution for the antiprotons.

%%%%%%%%%%%%%%%%%%%%%%%%%%%%%%%%%%%%%% fig:feedplot

%%%%%%%%%%%%%%%%%%%%%%%%%%%%%%%%%%%%%% EOFigure

%%%%%%%%%%%%%%%%%%%%%%%%%%%%%%%%%%%%%%%%%%%%%%%%%%%%%%%%%%%%%%%%%%%
\subsection {Systematic Uncertainties}
%%%%%%%%%%%%%%%%%%%%%%%%%%%%%%%%%%%%%%%%%%%%%%%%%%%%%%%%%%%%%%%%%%%

%%%%%%%%%%%%%%%%%%%%%%%%%%%%%%%%%%%%%%Table II 
\begin{table*}[tbh]
\caption{\label{tab:errors}
Systematic errors in percent on particle yields in d+Au and p+p
collisions. These values are independent of $p_T$.}
\begin{ruledtabular}
\begin{tabular}{lcc}
    & d+Au & p+p \\\hline
geometric acceptance correction & 4 & 4 \\
track matching & 9 & 8 \\
timing variations & 5 & 5 \\
reconstruction efficiency correction & 3 & 4 \\
energy loss correction & 1 & 2 \\
trigger bias & - & 4 \\\hline
\end{tabular}
\end{ruledtabular}
\end{table*}

%%new intro to systematic errors BVJ
%%

Systematic uncertainties on the hadron spectra are estimated as in
reference \cite{ppg026}. Various sets of $p_T$ spectra and ratios
of different particle types were made by varying the cut parameters
such as fiducial cut to check acceptance corrections, 
track association (i.e. matching) windows, and PID cuts,
from those used in the analysis. For each of these spectra
and ratios the same changes in cuts were made in the Monte Carlo 
analysis. The uncertainties were evaluated by comparing fully corrected
spectra and ratios from different cuts. 
The resulting uncertainties from each cut are
given in Table II and added in quadrature to yield overall systematic
uncertainties. 

%% rearranged from draft 0, for better flow  BVJ
%%  and slightly expanded to include each entry in table II.
%%

Additional systematic uncertainties on the hadron spectra arise from 
time variations in the TOF timing (slat-by-slat and run-by-run variations), the small
remaining contamination by other species after matching and PID 
cuts, uncertainty in the corrections for track reconstruction 
efficiency, and uncertainty on particle energy loss in the detector 
material. These uncertainties, which do not depend measurably on the 
hadron momentum, are listed in Table II. The sizable uncertainty 
in the matching cut is due to the non-Gaussian tails on the z-coordinate matching
distributions. The track matching in this direction is limited by a relatively poor 
vertex resolution determined by the BBC in p+p and d+Au collisions due to the small 
multiplicities. The quoted 3\%-4\% uncertainty in the reconstruction efficiency
correction represents the maximum local discrepancy between efficiencies
measured with strict and loose track quality cuts.

Uncertainties
due to particle identification cuts are momentum dependent. For
protons and antiprotons, the identification uncertainty is 8\% 
at low $p_T$ and decreases to 3\% at high $p_T$. Kaons at low
momentum have 10\% PID uncertainty, decreasing to 3\% at high $p_T$.
For pions the uncertainty increases from 4\% to 10 \% with increasing
$p_T$. Kaon and proton uncertainties decrease with increasing $p_T$
because energy loss and decay corrections become smaller. The
pion uncertainties are dominated by the particle identification
performance, which worsens with increasing $p_T$.

%$p_T$. 
%For protons and antiprotons there is an additional systematic
%error arising from the feed down correction. This is of the order
%of X\% for both protons and antiprotons, in both p+p and d+Au.

The systematic error on the feed-down
proton spectrum is 26\%, primarily due to uncertainty in the 
measured $\Lambda$ spectra and particle composition.
The resulting systematic error on the final prompt proton and antiproton
spectra is of the order of 10\% in both p+p and d+Au.
The systematic error on the proton to pion ratio is 12$\%$, 
including the uncertainty on $\overline\Lambda/\Lambda$.

Systematic uncertainties on the d+Au nuclear modification factors mostly cancel
as the p+p and d+Au data were collected immediately following one another,
and detector performance was very similar.
The overall systematic error in the nuclear modification factor is 
due to uncertainties in the reconstruction efficiencies, fiducial
volumes, and small run-by-run variations. It is approximately
10\%, independent of particle species and \pt. An additional d+Au scale uncertainty 
is shown as boxes around 1.0 in the figures; this 
is the quadrature sum of uncertainties on the p+p cross section of 9.6\%, 
and the number of binary collisions in the each centrality bin (presented 
in Table I).

The systematic error on the Au+Au nuclear modification factors is
derived by propagating the systematic errors
on p+p and Au+Au data \cite{ppg026} to the final ratio. The
average systematic error for pions is approximately 15\%, while
for protons and antiprotons it is on the order of  19\%.
The normalization uncertainty, as in d+Au, is the quadrature sum of 
uncertainties on the p+p cross section and the error on the number of 
binary collisions in the corresponding Au+Au centrality bin from
reference \cite{ppg026}. We
note that for the most central Au+Au bin (0-5\%), $N_{coll}$=1065.4
and the uncertainty is $\pm$105.3; 
in the most peripheral bin (60-92\%), 
 $N_{coll}$=14.5 $\pm$ 4.0 .

%%%%%%%%%%%%%%%%%%%%%%%%%%%%%%%%%%%%%%%%%%%%%%%%%%%%%%%%%%%%%%%%%%%%%%%%
\section { Results}

\subsection {Hadron spectra}

%%%%%%%%%%%%%%%%%%%%%%%%%%%%%%%%%%%%%%%% fig:pions
%\vspace{-3cm}
\begin{figure}
\begin{center}
\includegraphics[width=1.0\linewidth]{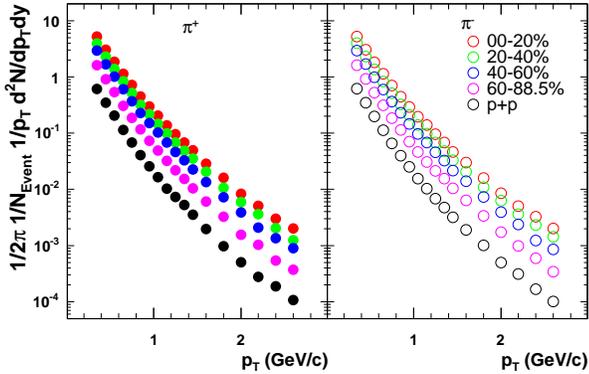}
\caption{(color online) Invariant yields at mid-rapidity for positive and
negative pions as a function of $p_T$ for 
various centrality classes in d+Au and p+p collisions. 
The error bars show statistical uncertainties only 
and are typically smaller than the data points. 
}
\label{fig:pions} 
\end{center}
\end{figure} 

%%%%%%%%%%%%%%%%%%%%%%%%%%%%%%%%%%%%%%%% fig:kaons
%\vspace{-3cm}
\begin{figure}
\begin{center}
\includegraphics[width=1.0\linewidth]{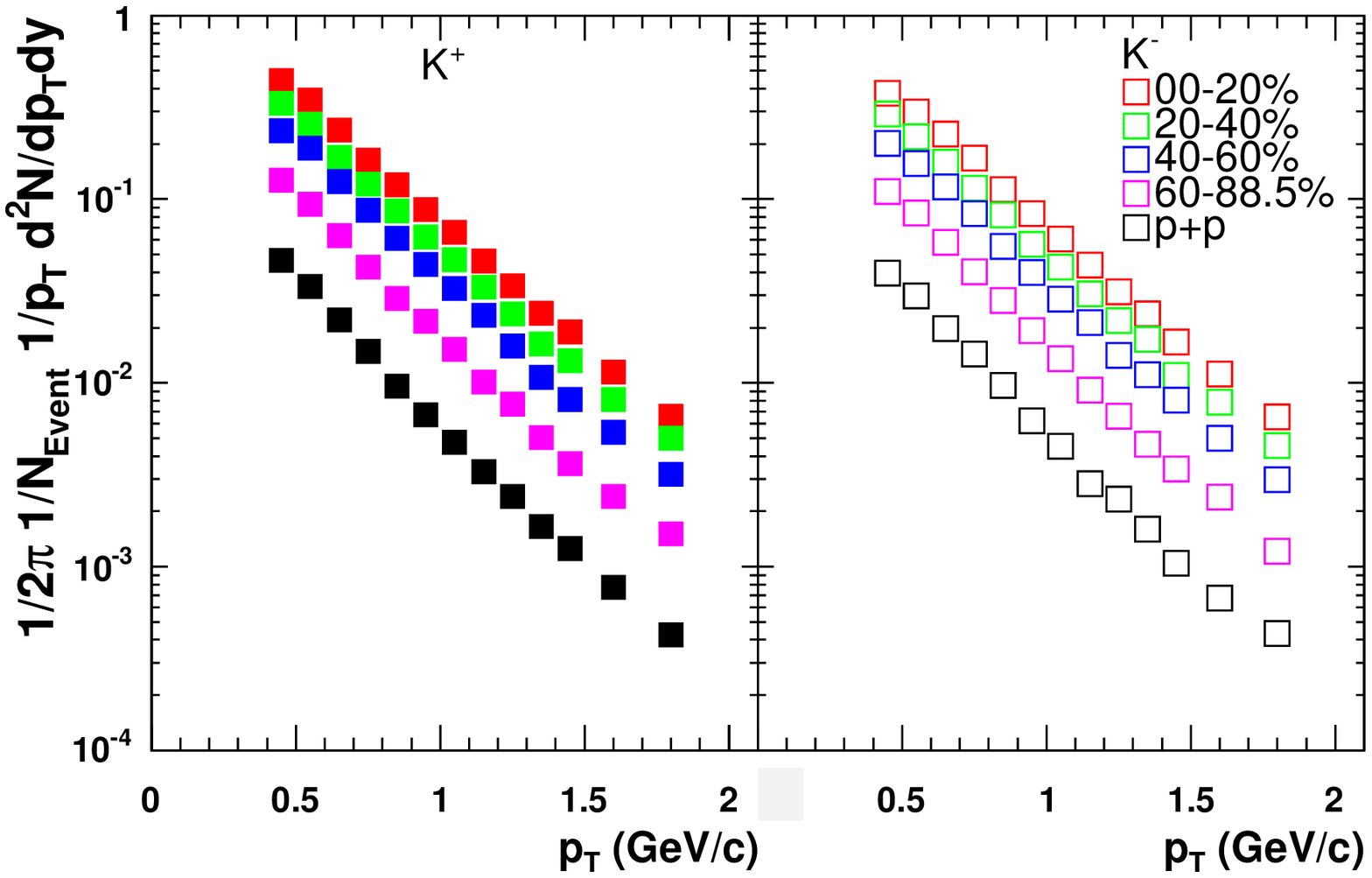}
\caption{(color online) Invariant yields at mid-rapidity for positive and
negative kaons 
as a function of $p_T$ for 
various centrality classes in d+Au and p+p collisions. 
The error bars show statistical uncertainties only
and are typically smaller than the data points. 
}
\label{fig:kaons} 
\end{center}
\end{figure} 

%%%%%%%%%%%%%%%%%%%%%%%%%%%%%%%%%%%%%%%% fig:protons
%\vspace{-3cm}
\begin{figure}
\begin{center}
\includegraphics[width=1.0\linewidth]{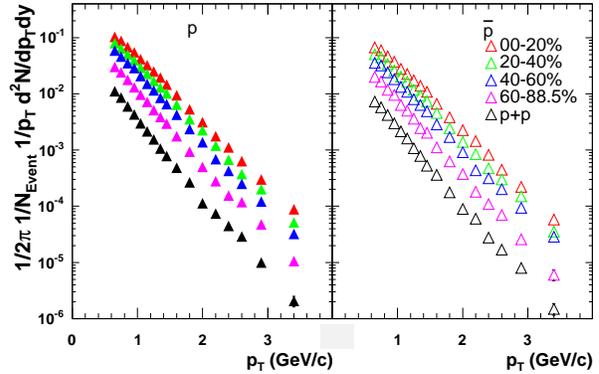}
\caption{(color online) Invariant yields at mid-rapidity for protons and
antiprotons as a function of $p_T$ for 
various centrality classes in d+Au and p+p collisions. 
The error bars show statistical uncertainties only
and are typically smaller than the data points.  
}
\label{fig:protons} 
\end{center}
\end{figure} 

The fully corrected $p_T$ distributions of $\pi,~K,~p,~\rm{and}
~\overline{p}$ for the four d+Au centrality bins and for p+p
collisions are shown
in Figures~\ref{fig:pions}, \ref{fig:kaons}, and \ref{fig:protons},
respectively. Pions show a power law spectral shape, while kaons 
and protons are exponential.

\begin{figure}
\begin{center}
\includegraphics[width=1.0\linewidth]{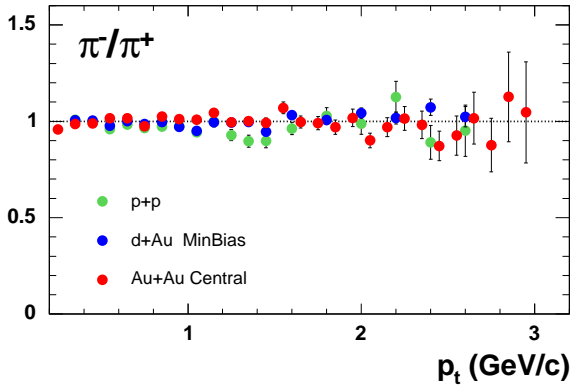}
\caption{(color online) Ratio of mid-rapidity spectra for $\pi^-$ to
$\pi^+$ in d+Au, p+p and central Au+Au collisions. 
The error bars show statistical uncertainties only. 
}
\label{fig:piminusplus} 
\end{center}
\end{figure}

\begin{figure}
\begin{center}
\includegraphics[width=1.0\linewidth]{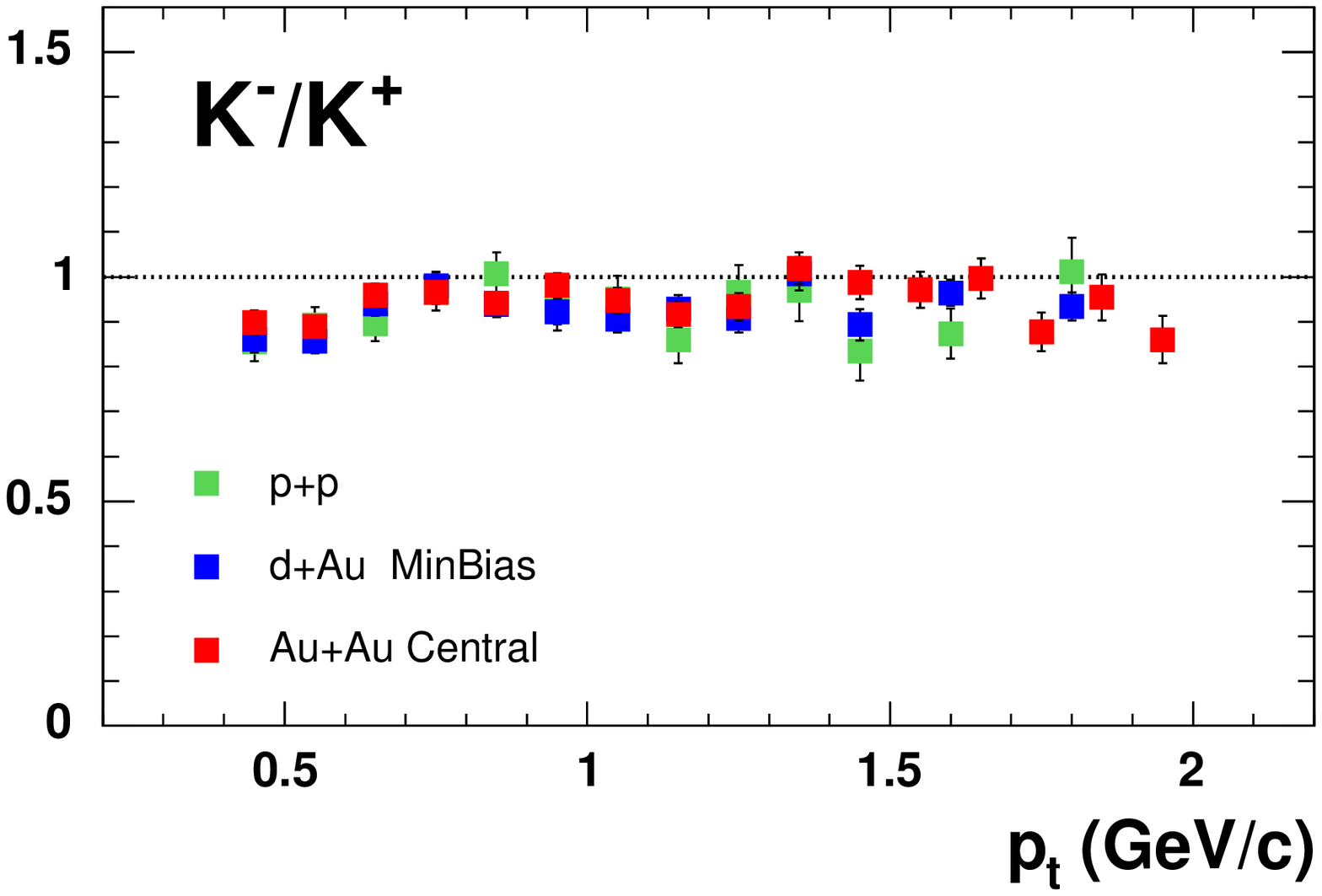}
\caption{(color online) Ratio of mid-rapidity spectra for $K^-$ to
$K^+$ in d+Au, p+p and central Au+Au collisions. 
The error bars show statistical uncertainties only. 
}
\label{fig:kminusplus} 
\end{center}
\end{figure}

\begin{figure}
\begin{center}
\includegraphics[width=1.0\linewidth]{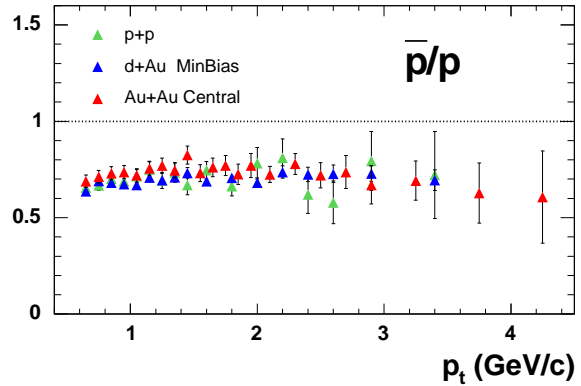}
\caption{(color online) Ratio of mid-rapidity spectra for antiprotons to
protons in d+Au, p+p and central Au+Au collisions \cite{ppg026}. 
The error bars show statistical uncertainties only. 
}
\label{fig:pbarp} 
\end{center}
\end{figure}

In order to probe the hadron production mechanism, it is instructive
to compare particle and anti-particle spectra. Figures~\ref{fig:piminusplus},
\ref{fig:kminusplus}, and \ref{fig:pbarp} show the ratios of antiparticle
to particle production as a function of $p_T$ in p+p, d+Au and, for
comparison, central Au+Au collisions from reference \cite{ppg026} for
$\pi, K$ and p, respectively.
For all three hadron species the ratios are flat with $p_T$. d+Au
yield ratios are in good agreement with p+p collisions, and the
ratios remain the same even in central Au+Au collisions. The
production ratio of antiparticle to particle is 
0.99 $\pm$ 0.01(stat) $\pm$ 0.06(syst) for pions  
and 0.92 $\pm$ 0.01(stat) $\pm$ 0.07(syst) for kaons
in both minimum bias d+Au and p+p collisions. The 
antiproton to proton ratio is measured to be
0.70 $\pm$ 0.01(stat) $\pm$ 0.08(syst) in minimum bias d+Au collisions
and 0.71 $\pm$ 0.01(stat) $\pm$ 0.08(syst) in p+p collisions.
All ratios are consistent within errors with values reported by PHOBOS \cite{PHOBOS}
and BRAHMS \cite{BRAHMS}.

\subsection {Nuclear Modification Factors}

%%%%%%%%%%%%%%%%%%%%%%%%%%%%%%%%%%%%%%%% fig:piminusplus
%\vspace{-3cm}

%%%%%%%%%%%%%%%%%%%%%%%%%%%%%%%%%%%%%%%% fig:kminusplus
%\vspace{-3cm}

%%%%%%%%%%%%%%%%%%%%%%%%%%%%%%%%%%%%%%%% fig:pbarp
%\vspace{-3cm}

The measurement of identified hadrons in both d+Au and p+p
collisions allows study of the centrality dependence of
the nuclear modification factor in d+Au.
A standard way to quantify nuclear medium effects on high $p_T$ 
particle production in nucleus-nucleus collisions is provided by the 
{\it nuclear modification factor}. This is the ratio of the d+A 
invariant yields to the binary collision scaled p+p invariant yields:
\begin{equation} 
%R_{dA}(p_T)\,=\,\frac{(1/N^{evt}_{dA})\,d^2N_{dA}/dy dp_T}{\langle %N_{coll}\rangle \, 
%(1/N^{evt}_{pp})\,d^2N^_{pp}/dy dp_T},
R_{dA}(p_T)\,=\,\frac{(1/N^{evt}_{dA})\,d^2N_{dA}/dy dp_T}{T_{dAu}\, 
d^2\sigma^{pp}_{inel}/dy dp_T},
\label{eq:R_dA}
\end{equation}
where $T_{dAu} = \langle N_{coll}\rangle/\sigma^{pp}_{inel}$ describes the
nuclear geometry, and $d^2\sigma^{pp}_{inel}/dy dp_T$ for p+p collisions
is derived from the measured p+p cross section. 
$\langle N_{coll}\rangle$ is the average number of 
inelastic nucleon-nucleon collisions determined from simulation using 
the Glauber model as input, as described in section~\ref{subsec:exp}. $N^{evt}_{dA}$ is the number of d+Au events
in the relevant centrality class.

%%%%%%%%%%%%%%%%%%%%%%%%%%%%%%%%%%%%%%%% fig:rdauminbias
%\vspace{-0.5cm}
\begin{figure}
%\vspace{-0.9cm}
\includegraphics[width=1.0\linewidth]{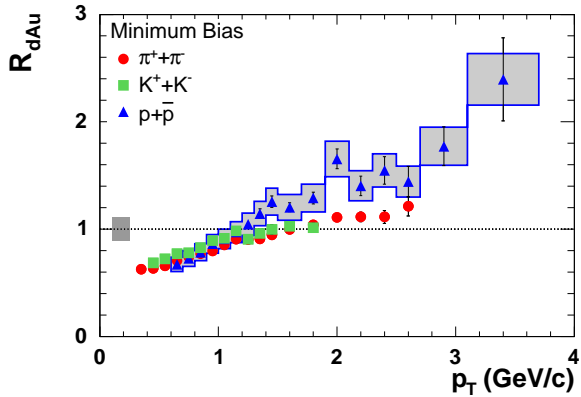}
%\vspace{0.cm}
\caption{(color online) Nuclear modification factor $R_{dA}$  for pions, kaons 
and protons in d+Au collisions for minimum bias events.
The error bars represent the statistical errors.
The box around 1.0 shows uncertainties in the p+p absolute cross section
and in the calculation of $N_{coll}$. For the proton and antiproton
$R_{dA}$, the $\sim$10$\%$ systematic uncertainty is also presented as boxes around the
points. The systematic uncertainty on the pion and kaon $R_{dA}$ is similar but not shown in the
picture for clarity.
}
\label{fig:rdauminbias}
\end{figure} 

Figure~\ref{fig:rdauminbias} shows $R_{dA}$ for pions, kaons and
protons for minimum bias d+Au collisions. We observe a 
nuclear enhancement in the production of hadrons with $p_T \ge$ 
1.5 - 2 GeV/c in d+Au collisions, compared to that in p+p. 
As was already suggested when comparing the enhancement for
inclusive charged hadrons with that of neutral pions \cite{ppg028},
there is a species dependence in the Cronin effect. The
Cronin effect for charged pions is small, as
was observed for neutral pions. The nuclear enhancement for
protons and antiprotons is considerably larger. The kaon 
measurement has a more limited kinematic range, but the
%magnitude of the enhancement 
$R_{dA}$ is in agreement with that of
the pions at comparable $p_T$.

%%%%%%%%%%%%%%%%%%%%%%%%%%%%%%%%%%% fig:rdau
%\vspace{-0.5cm}
\begin{figure*}[thb]
%\vspace{-0.9cm}
\includegraphics[width=0.6\linewidth]{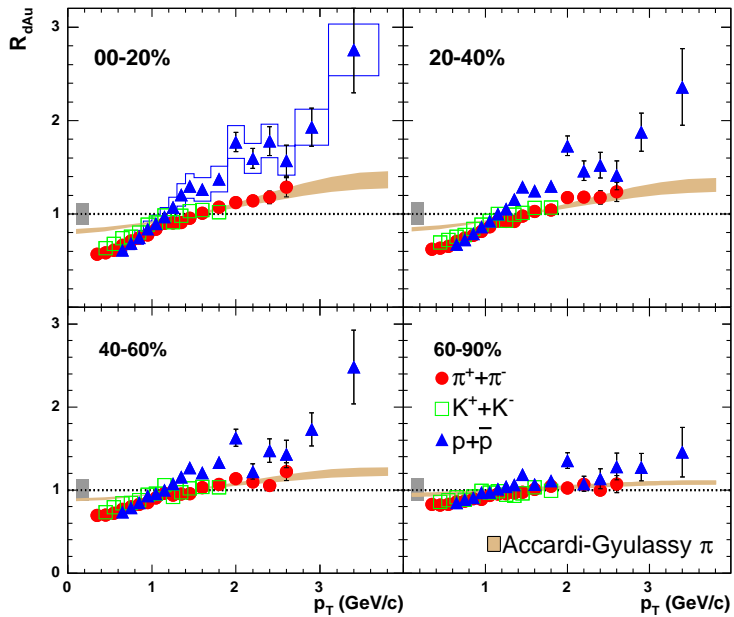}
%\vspace{0.cm}
\caption{(color online) Nuclear modification factor $R_{dA}$  for pions, kaons 
and protons in d+Au collisions in four centrality bins.
The error bars represent the statistical errors.
Boxes around 1.0 show uncertainties in the p+p absolute cross section
and in the calculation of $N_{coll}$. 
For the proton and antiproton $R_{dA}$, the $\sim$10$\%$ systematic 
uncertainty is also presented as boxes around the points.
The systematic uncertainty on the pion and kaon $R_{dA}$ is similar but not shown in the
picture for clarity.
Solid bands show the calculation of the
nuclear modification factors for pions by Accardi and Gyulassy \cite{Accardi}.
}
\label{fig:rdau}
\end{figure*} 

Figure~\ref{fig:rdau} shows $R_{dA}$ for pions, kaons and
protons in the four d+Au centrality bins.
%The nuclear modification factor for kaons appears to
%saturate near 1, but 
Peripheral d+Au collisions ($ \langle N_{coll} \rangle = 3.1\pm0.3$) 
do not show any modification of
high momentum hadron production, compared to that in p+p collisions.
At $p_T \le $ 1 GeV/c, the nuclear modification factor falls below
1.0. This is to be expected as soft particle production scales 
with the number of participating nucleons, not with the number 
of binary nucleon-nucleon collisions. More central collisions 
show increasing nuclear enhancement in both high $p_T$ pion 
and proton production.

The bands in Fig.~\ref{fig:rdau} 
show a calculation of the Cronin effect for pions by Accardi
and Gyulassy, using a pQCD model of multiple semi-hard collisions
and taking geometrical shadowing into account \cite{Accardi}. 
The agreement above 1 GeV/c, where the calculation should be
reliable, is very good for all four centrality bins. 
This agreement illustrates that the multiple partonic scattering and
nuclear shadowing alone can explain the observed Cronin effect
%the effects of multiple partonic scattering and nuclear shadowing. 
%The quantitative agreement
and leaves very little room for gluon saturation effects
%dynamical shadowing effects 
in the 
nuclear initial state at mid-rapidity at RHIC \cite{Accardi}.

\subsection {Centrality Dependence}

\begin{figure}[bht]
\includegraphics[width=1.0\linewidth]{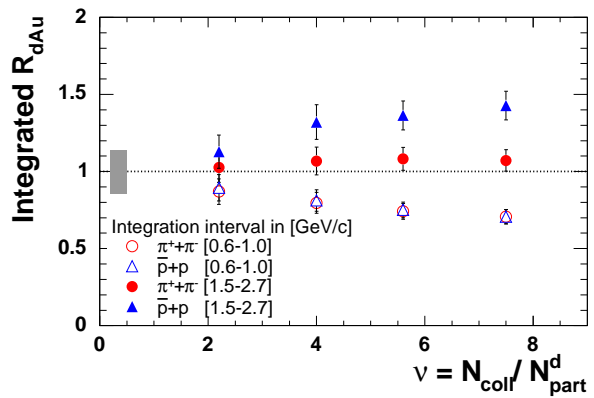}
\caption{(color online) Integrated $R_{dA}$ for pions and protons in two momentum bins as a function of the 
number of collisions suffered by 
the deuteron participant $\nu$. Error bars indicate the quadrature sum
of statistical errors and uncertainties on the number of collisions
bin-by-bin. The solid box on the left shows the magnitude
of the centrality independent uncertainties. }
\label{fig:nu} 
\end{figure}

We further probe the effect of cold nuclear matter upon the hadron
production using the number of collisions suffered by each 
projectile nucleon for the four centrality bins.
Figure~\ref{fig:nu} compares the centrality dependence of
$R_{dAu}$ for pions
and protons in two momentum bins. The modification factors are plotted
as a function of $\nu = N_{coll}/N_{part}^d$, the number of collisions 
per participating deuteron
nucleon. The lower momentum bin, for 0.6 $ \le p_T
\le$ 1.0 GeV/c, is chosen in the region where $R_{dAu}$ is less than 1.0, 
and hadron yields scale very nearly with the number of nucleons
participating in the collision, rather than with the number 
of binary collisions. 
As expected, $R_{dAu}$ decreases with $\nu$
in this $p_T$ range, with negligible difference between pions and
protons.
In the higher $p_T$ bin, $R_{dAu}$ increases with the number of 
collisions, with a notably larger rate of increase 
for baryons than for mesons.
Though the $R_{dAu}$ values for higher $p_T$ hadrons appear to
flatten with increasing centrality, the uncertainties are too
large to allow a definitive conclusion about saturation with
the number of collisions suffered by each participant nucleon 
\cite{PappLevai}.

%%%%%%%%%%%%%%%%%%%%%%%%%%%%%%%%%%%%%%%% fig:nu
%\vspace{-0.5cm}

\clearpage

\subsection {Cronin Effect For Baryons}

%%%%%%%%%%%%%%%%%%%%%%%%%%%%% fig:ratiordau
%\vspace{-0.5cm}
\begin{figure}
\includegraphics[width=1.0\linewidth]{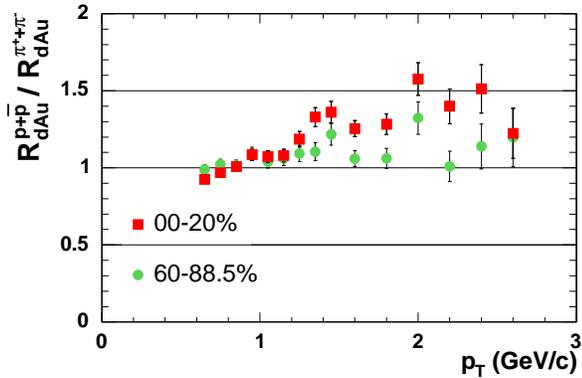}
\caption{(color online) Ratio of the proton over the pion nuclear 
modification in d+Au collisions, for central and
peripheral events. Error bars indicate statistical 
errors only.}
\label{fig:ratiordau} 
\end{figure} 

Figure~\ref{fig:ratiordau} shows the ratio of the nuclear
modification factors observed for protons and antiprotons to
that for pions in the most central and the most peripheral d+Au
collisions. The enhancement for protons is stronger, by 
approximately 30\%-50\% in the most central collisions; this 
was also reported by STAR \cite{starpiddau}. 
The increasing difference between baryons and mesons for
more central collisions indicates that the baryon production 
mechanism appears to depend upon the surrounding nuclear 
medium already in d+Au collisions. We note, however, that 
the species dependence of the Cronin effect in d+Au collisions
is much smaller than the factor of $\approx$3 enhancement of 
protons in central Au+Au collisions, as can be seen by comparing 
the nuclear modification factors for pions and protons in 
central Au+Au collisions shown below in section~\ref{subsec:comp_to_au+au}.

%%%%%%%%%%%%%%%%%%%%%%%%%%%%% fig:lowCronin
%\vspace{-0.5cm}
\begin{figure}
\includegraphics[width=1.0\linewidth]{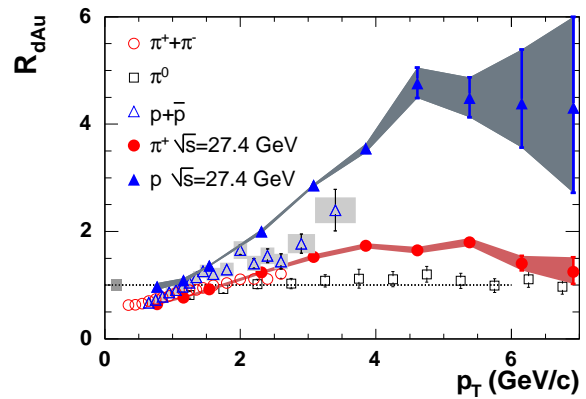}
\caption{Nuclear modification factors for charged pions, 
neutral pions \cite{ppg028}, and protons and antiprotons from 
minimum bias d+Au from this work 
(shown by open symbols) compared to nuclear modification
factors calculated from the per-beam nucleon cross sections
reported for $\sqrt{s}$ = 27.4 GeV p + A collisions 
\cite{antreasyan} (closed symbols and continuous bands)}.
\label{fig:lowCronin} 
\end{figure} 

By converting the nuclear modification factors to  
per-beam-nucleon cross section 
ratios and vice versa, it is possible to compare to measurements of the Cronin
effect at other energies.
 %It is also possible to convert
%in the other direction. 
Figure~\ref{fig:lowCronin} shows
nuclear modification factors from this work compared to
those derived from $\alpha$-factors measured 
at lower energies \cite{antreasyan}, in a similar manner 
to that used in \cite{straub} to calculate per-nucleon cross
section ratios. In order to make the transition to nuclear
modification factors from per-nucleon cross section ratios
we assume that $\sigma_{d+A}=2 \times \sigma_{p+A}$ and
$\sigma_{p+d}=2 \times \sigma_{p+p}$ at the low energy of interest, 
which is a very reasonable approximation for all particle species \cite{antreasyan}.
%The atomic mass numbers used for the conversions are
%  $A_D=2.014$, $A_{Be} = 9.012$, $A_W = 183.840$, and $A_{Au} = 196.966$.

The observed species dependence of the enhancement
is similar to that measured in lower energy collisions \cite{straub}. 
The magnitude of the enhancement for pions at $p_T > 3$ GeV/c is
%somewhat 
larger at $\sqrt{s}$ = 27.4 GeV than at 200 GeV. 
Protons and antiprotons are also more 
enhanced at the lower beam 
energy: a factor of 3.5 at $p_T \approx$ 4 GeV, as compared with 
a factor of 2 at $\sqrt{s}$ = 200 GeV.
This energy dependence of the Cronin effect for pions
has been interpreted as evidence for a different
production mechanism for high $p_T$ hadrons at RHIC compared to
lower energies \cite{boris}. In this model, high $p_T$ hadrons 
are produced
incoherently on different nucleons at low energy, while in 
higher energy collisions the production amplitudes can interfere
because the process of gluon radiation is long compared to the
binary collision time. Coherent radiation from different
nucleons is subject to Landau-Pomeranchuk suppression. However,
the difference between baryon and meson Cronin effect is not
predicted by this model.

\section {Discussion}

Traditional explanations of the Cronin effect all
involve multiple scattering of incoming partons that lead  
to an enhancement at intermediate $p_{T}$ \cite{LevPetersson}. 
There are various theoretical models of the multiple scattering, 
which predict
somewhat different dependence upon the number of scattering centers.
The observed centrality or $\nu$ dependence for pions is well-reproduced 
by semi-hard initial state scattering \cite{Accardi}
as shown in Fig.~\ref{fig:rdau}; see also 
\cite{PappLevai, Vitev, Wang}. The models include 
initial state multiple scattering as well as geometrical shadowing. 
However, none of these models would predict a species dependent 
Cronin effect, 
as initial state parton scattering precedes fragmentation
into the different hadronic species. The markedly larger Cronin
effect for protons and antiprotons requires 
processes in addition to initial state multiple scattering in baryon
production at moderate transverse momenta.

Recently, Hwa and collaborators \cite{Hwa} have shown 
an alternative explanation of the Cronin effect, attributed to 
the recombination of shower quarks with those from the medium
in d+Au  collisions.  Such models do predict a larger Cronin
effect for protons than pions, and may be justified by the rather
short formation times of $p_T$ = 2 - 4 GeV/c protons. 
According to the uncertainty principle, 
the formation time in the rest frame of the hadron can be
related to the hadron size $R_h$. In the laboratory frame,
the formation time of a hadron with mass $m_h$ and energy 
$E_h$ is given by
$\tau_f \approx R_h {{E_h}\over{m_h}}$. 
For a $p_T$ = 2.5 GeV/c pion, the formation
time is 9-18 fm/c (for $R_h$ = 0.5 - 1.0 fm), well outside
the collision region. However, for $p_T$ = 2.5 GeV/c protons,
the corresponding formation time is only 2.7 fm/c, suggesting
that the hadronization process may well begin in or near the 
nuclear medium.

\subsection{Comparison to Au+Au Collisions}
\label{subsec:comp_to_au+au}

%%%%%%%%%%%%%%%%%%%%%%%%%%%%%%%%%%%%%%%% fig:ptopi
%\vspace{-0.5cm}
\begin{figure}
\includegraphics[width=1.0\linewidth]{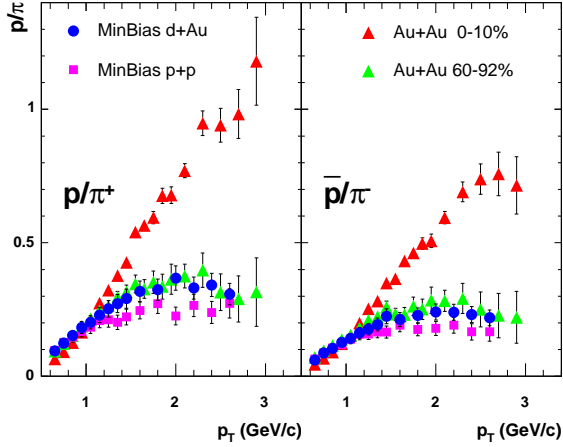}
\caption{(color online) The ratio of feed-down corrected
protons to $\pi^+$ and antiprotons to
$\pi^-$ in minimum bias p+p and d+Au compared to peripheral
and central Au+Au collisions. Statistical error bars are shown.}
\label{fig:ptopi} 
\end{figure} 

The proton to pion ratio from minimum bias p+p and minimum bias d+Au 
are compared to each other and to central and peripheral Au+Au 
collisions in Fig.~\ref{fig:ptopi}. As noted above,
protons and antiprotons are feed-down corrected in each system.

The $p/\pi$ ratio in d+Au is very similar to that in peripheral Au+Au 
collisions, and lies slightly above
the p+p ratio. The $p/\pi$ ratio in central 
Au+Au collisions is, however,  much larger. The difference between the 
ratio in d+Au and central Au+Au clearly indicates that baryon yield
enhancement is not simply an effect of sampling a large nucleus in the
initial state. The large enhancement requires the presence of a substantial 
volume of nuclear medium.
%with high energy density.

%%%%%%%%%%%%%%%%%%%%%%%%%%%%%%% fig:rauaudaupi
%\vspace{-0.5cm}
\begin{figure}
\includegraphics[width=1.0\linewidth]{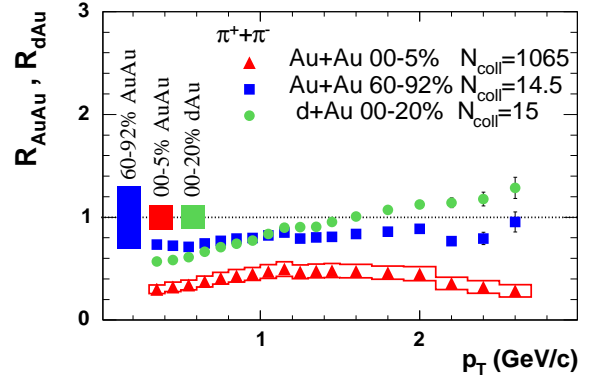}
\caption{(color online) Nuclear modification factors for pions,
comparing central and peripheral Au+Au collisions to central d+Au.
It should be noted that the number of binary nucleon-nucleon 
collisions in peripheral Au+Au and central d+Au is very similar.
Solid bars on the left indicate normalization uncertainties
in the p+p absolute cross section and in the calculation of $N_{coll}$
for the three systems. Error bars indicate statistical errors only,
while for the most central Au+Au case the systematic errors, discussed 
in the text, are shown as boxes around the points.
}
\label{fig:rauaudaupi} 
\end{figure} 

%%%%%%%%%%%%%%%%%%%%%%%%%%%%%%%% fig:rauaudauk
%\vspace{-0.5cm}

\begin{figure}
\includegraphics[width=1.0\linewidth]{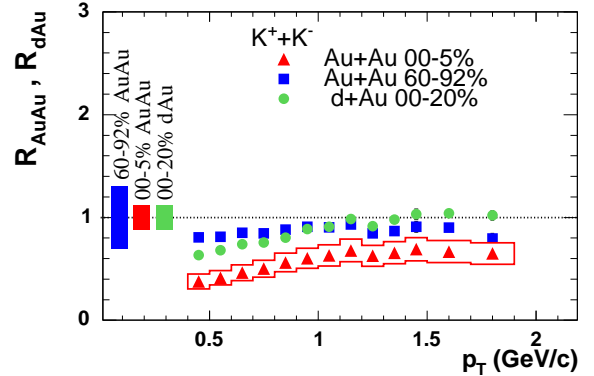}
\caption{(color online) Nuclear modification factors for kaons,
comparing central and peripheral Au+Au collisions to central d+Au.
Statistical and systematic errors are represented 
as in Fig.~\protect\ref{fig:rauaudaupi}.
}
\label{fig:rauaudauk} 
\end{figure} 

%%%%%%%%%%%%%%%%%%%%%%%%%%%%%%%% fig:rauaudaup
%\vspace{-0.5cm}
\begin{figure}
\includegraphics[width=1.0\linewidth]{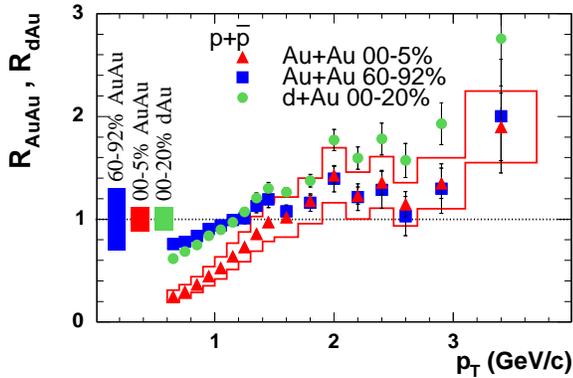}
\caption{(color online) Nuclear modification factors for protons and antiprotons,
comparing central and peripheral Au+Au collisions to central d+Au.
Statistical and systematic errors are represented 
as in Fig.~\protect\ref{fig:rauaudaupi}.
}
\label{fig:rauaudaup} 
\end{figure} 

Figures~\ref{fig:rauaudaupi}, \ref{fig:rauaudauk} and \ref{fig:rauaudaup}
 compare the nuclear modification factors for pions, kaons and 
(anti)protons in Au+Au and d+Au collisions. 
The p+p data from this work allow, for the first time, 
the calculation 
of nuclear modification factors in Au+Au by PHENIX. 
The Au+Au data are taken from reference \cite{ppg026}. 
It should be noted that common fluctuations between $R_{dAu}$ and
$R_{AuAu}$ in Figures~\ref{fig:rauaudauk} and \ref{fig:rauaudaup}
arise because the p+p denominators are common.

Central and
peripheral Au+Au collisions are compared to central d+Au collisions,
which have a similar number of binary collisions as the peripheral
Au+Au sample. Pions show a much lower $R_{AA}$ at high $p_T$ in central
than in peripheral Au+Au collisions, as expected from the large
energy loss suffered by the partons in central collisions.
The nuclear modification factor rises faster with $p_T$ 
in d+Au than in peripheral Au+Au, despite the comparable
number of binary collisions. As Au+Au involves a second Au nucleus,
shadowing effects can be expected to be larger,
reducing the observed Cronin effect. 

The proton and antiproton nuclear modification factors
show a quite different trend. The Cronin effect, 
larger than 1.0
at higher $p_T$ values, is independent of centrality in Au+Au
collisions. This feature was already observed as binary
collision scaling
of proton and antiproton production in the central/peripheral
collision yield ratios \cite{ppg015}. The Cronin effect in d+Au
is at least as large as in peripheral Au+Au. The difference
indicates that baryon production must involve a complex
interplay of processes in addition to initial state 
nucleon-nucleon collisions.

%%%%%%%%%%%%%%%%%%%%%%%%%%%%%%%%%%%%%%%%%%%%%%%%%%%%%%%%%%%%%%%%%
\section {Conclusion}
%
%\marginpar{{\small \em Concl}}

We have presented the centrality and species dependence of
identified particle spectra in d+Au collisions, including the dependence
of the nuclear modification factor upon the number of collisions 
per participant nucleon. We also presented the first measurement
of $R_{AA}$ for pions, kaons, protons and antiprotons in Au+Au collisions.

The Cronin effect for charged pions is small, but non-zero.
The proton to pion ratio in d+Au is similar to that in
peripheral Au+Au,
while the corresponding ratio in p+p is somewhat lower.
The nuclear modification factor in d+Au for protons shows a larger 
Cronin effect than that for pions, and the difference increases
with collision centrality. This difference was seen, but never
fully understood, in lower energy collisions, and 
is not large enough
to account for the abundance of protons in central Au+Au collisions.
The difference between pions and protons does, however, indicate 
that the Cronin effect is
not simply due to multiple scattering of the incoming partons.
 $R_{AA}$ for protons and antiprotons confirms previous
observations that the production of high $p_T$ baryons in Au+Au
scales with the number of binary nucleon-nucleon collisions,
but the baryon yield per collision in Au+Au exceeds that in p+p. 
%Thus there 
%are two baryon mysteries, which likely both arise from interactions
%with the surrounding media as the baryons are formed.
%%%%%%%%%%%%%%%%%%%%%%%%%%%%%%%%%%%%%%%%%%%%%%%%%%%%%%%%%%%%
% Acknowledgements
%

%\section{Acknowledgements}   % Run-3 long from for PRC, PLB, etc.
\section*{Acknowledgements}

We thank the staff of the Collider-Accelerator and Physics
Departments at Brookhaven National Laboratory and the staff of
the other PHENIX participating institutions for their vital
contributions.  We acknowledge support from the Department of
Energy, Office of Science, Office of Nuclear Physics, the
National Science Foundation, Abilene Christian University
Research Council, Research Foundation of SUNY, and Dean of the
College of Arts and Sciences, Vanderbilt University (U.S.A),
Ministry of Education, Culture, Sports, Science, and Technology
and the Japan Society for the Promotion of Science (Japan),
Conselho Nacional de Desenvolvimento Cient\'{\i}fico e
Tecnol{\'o}gico and Funda\c c{\~a}o de Amparo {\`a} Pesquisa do
Estado de S{\~a}o Paulo (Brazil),
Natural Science Foundation of China (People's Republic of China),
Centre National de la Recherche Scientifique, Commissariat
{\`a} l'{\'E}nergie Atomique, and Institut National de Physique
Nucl{\'e}aire et de Physique des Particules (France),
Ministry of Industry, Science and Tekhnologies,
Bundesministerium f\"ur Bildung und Forschung, Deutscher
Akademischer Austausch Dienst, and Alexander von Humboldt Stiftung (Germany),
Hungarian National Science Fund, OTKA (Hungary), 
Department of Atomic Energy (India), 
Israel Science Foundation (Israel), 
Korea Research Foundation, Center for High
Energy Physics, and Korea Science and Engineering Foundation (Korea),
Ministry of Education and Science, Rassia Academy of Sciences,
Federal Agency of Atomic Energy (Russia),
VR and the Wallenberg Foundation (Sweden), 
the U.S. Civilian Research and Development Foundation for the
Independent States of the Former Soviet Union, the US-Hungarian
NSF-OTKA-MTA, and the US-Israel Binational Science Foundation.

%REFERENCES:  Use \begin{references} and \end{references}.  Do not use
%             \begin{thebibliography} and \end{thebibliography}.
%             You may either 
%		(a) enter all citations explicitly or 
%               (b) use some "\def" shorthand notations.
%             Our first paper used approach (a) and our second used (b).
%             Here are the two reference lists as examples of how to proceed:

%%%%%%%\end{multicols}


\begin{thebibliography}{99}

\def\Journal#1#2#3#4{{#1}{\bf #2}, #3 (#4)}
\def\IJMPA{{Int. J. Mod. Phys.}~{\bf A}}
\def\EPJ{{Eur. Phys. J.}~{\bf C}}
\def\JPG{{J. Phys}~{\bf G}}
\def\JHEP{{J. High Energy Phys.}~}
\def\NCA{Nuovo Cimento~}
\def\NIM{Nucl. Instrum. Methods~}
\def\NIMA{{Nucl. Instrum. Methods}~{\bf A}}
\def\NPA{{Nucl. Phys.}~{\bf A}}
\def\NPB{{Nucl. Phys.}~{\bf B}}
\def\PLB{{Phys. Lett.}~{\bf B}}
\def\PLC{Phys. Repts.\ }
\def\PRL{Phys. Rev. Lett.\ }
\def\PRL{Phys. Rev. Lett.\ }

%\def\PRD{{Phys. Rev.}~{\bf D}}
\def\PRD{{Phys. Rev. D}}

\def\PRC{{Phys. Rev.}~{\bf C}}
\def\ZPC{{Z. Phys.}~{\bf C}}

%1
\bibitem{cronin} J.W.~Cronin {\it et al.},  \Journal{\PRD}{11}
{3105} {1975}.
%2
\bibitem{antreasyan} D.~Antreasyan {\it et al.}, \Journal{\PRD}{19}{764}{1979}.
%3
\bibitem{straub} P.B.~Straub {\it et al.}, \Journal{\PRL}{68} {452}{1992}.
%4
\bibitem{ppg003} K.~Adcox {\it et al.},  [PHENIX Collaboration],
			\PRL{\bf 88}, 022301 (2002).
%5
\bibitem{ppg006} K.~Adcox {\it et al.},  [PHENIX Collaboration],
			\PRL{\bf 88}, 242301 (2002).
%6
\bibitem{ppg015} S.S.~Adler {\it et al.} [PHENIX Collaboration], 
\PRL{\bf 91}, 172301 (2003).

%7
\bibitem{ppg026} S.S. Adler {\it et al.}, [PHENIX Collaboration].
			\PRC{\bf 69}, 034909 (2004).

%8
\bibitem{ISR} B.~Alper {\it et al.},
\NPB{\bf 100}, 237 (1975).

%9
\bibitem{recomb}
R.C. Hwa and C.B. Yang, \PRC{\bf 67}, 034902 (2003) and \PRC{\bf 70}, 037901
(2003);
R.J. Fries, B. Muller, C. Nonaka and S.A. Bass \PRL{\bf 90}, 202303 (2003) 
and \PRC{\bf 68}, 044902 (2003);
V. Greco, C.M. Ko and P. Levai, \PRL{\bf 90}, 202302 (2003) and
\PRC{\bf 68}, 034904 (2003).

%10
\bibitem{junct}
G. Rossi and G. Veneziano, \NPB{\bf 123}, 507 (1977);
D. Kharzeev, \PLB{\bf 378}, 238 (1996);
S.E. Vance, M. Gyulassy and X.-N. Wang, \PLB{\bf 443}, 45 {1998}.

%11
\bibitem{colorglas}
D. Kharzeev, E. Levin and L. McLerran, \PLB{\bf 561}, 93 (2003).

%12
\bibitem{LevPetersson} M. Lev and B. Petersson, \ZPC{\bf 21}, 155 (1983).

%13
\bibitem{Accardi} A. Accardi and M. Gyulassy, \PLB{\bf 586}, 244 (2004).

%14
\bibitem{PappLevai} G.~Papp, P.~Levai, G.~Fai, \PRC{\bf 61}, 021902(R) (1999).
%SATURATING CRONIN EFFECT IN ULTRARELATIVISTIC PROTON NUCLEUS COLLISIONS.
%e-Print Archive: nucl-th/9903012

%15
\bibitem{Vitev} I.~Vitev, M.~Gyulassy, \PRL{\bf89},  252301 (2002).
%HIGH P(T) TOMOGRAPHY OF D + AU AND AU+AU AT SPS, RHIC, AND LHC.
%e-Print Archive: hep-ph/0209161

%16
\bibitem{Wang} X.N.~Wang, \PRC{\bf 61}, 064910 (2000).
%SYSTEMATIC STUDY OF HIGH P(T) HADRON SPECTRA IN P P, P A AND A A COLLISIONS
% FROM SPS TO RHIC ENERGIES. %e-Print Archive: nucl-th/9812021

%17
\bibitem{Hwa} R.C. Hwa and C.B. Yang, \PRL{\bf 93}, 082302 (2004). 

%18
\bibitem{nim_phenix} K. Adcox {\it et al.}, [PHENIX Collaboration],
			\NIM{\bf A499}, 469 (2003). 
			In addition to this overview, detailed descriptions
			of individual subsystems are published in the same volume.

%%19
\bibitem{vanderMeer} K.A. Drees and Z. Xu, ``Proceedings of the PAC2001 Conference'' 3120 (2001). 
http://epaper.kek.jp/p03/PAPERS/TPPB032.PDF

%%20
\bibitem{GEANT} R. Brun, R. Hagelberg, N. Hansroul and J. Lassalle,
CERN-DD-78-2 (1978).

%21
%\bibitem{ppg024} S.S.~Adler {\it et al.}, [PHENIX Collaboration], 
%\PRL{\bf 91}, 072301 (2003).

%22
\bibitem{jpsi_paper} S.S.~Adler {\it et al.}, [PHENIX Collaboration], 
\PRL{\bf 96}, 012304 (2006).

%23
\bibitem{glauber} R.J.~Glauber and G.~Matthiae, \NPB{\bf 21}, 135  (1970).

%24
\bibitem{ppg036} S.S.~Adler {\it et al.} [PHENIX Collaboration], 
\PRL{\bf 94}, 082302  (2005).
%nucl-ex/041154 (2004).Phys.Rev.Lett.94:082302,2005 

%25
\bibitem{hulthen} L.~Hulth\`en and M.~Sagawara, Handbuch der Physik {\bf 39} (1957).

%26
\bibitem{trackingNIM} K. Adcox {\it et al.}, [PHENIX Collaboration],
			\NIM{\bf A482}, 491 (2002). 


%27
\bibitem{UA5} 
R.E.~Ansorge {\it et al.}, [UA5 Collaboration], 
Nucl. Phys. {\bf B328}, 36 (1989).

%28
\bibitem{heinzpaper}
J.~Adams and M.~Heinz, [STAR Collaboration], nucl-ex/0403020, (2004).

%29
\bibitem{heinzppt}
M.~Heinz, [STAR Collaboration], J.\ Phys.\ G {\bf 31}, S141 (2005).
% Hot Quarks 2004, July 18-24, 2004 , Taos Valley, New Mexico. 

%30
\bibitem{cai}
X.~Z.~Cai, [STAR Collaboration], J.\ Phys.\ G {\bf 31}, S1015 (2005).
%X.~Cai [STAR Collaboration],
%8th International Conference on Strangeness in Quark Matter, 
%Cape Town, South Africa (2004).

%31
\bibitem{mtscaling}
R.~Witt, [STAR Collaboration], nucl-ex/0403021 (2004).

%32
\bibitem{PHOBOS}
B.B.~Back {\it et al.}, [PHOBOS Collaboration], \PRC{\bf 70}, 011901 (2004); \PRC{\bf 71}, 021901 (2005).

%8) CENTRALITY DEPENDENCE OF CHARGED ANTI-PARTICLE TO PARTICLE RATIOS NEAR MID RAPIDITY IN D + AU COLLISIONS AT S(NN)**(1/2) = 200-GEV.
%By PHOBOS Collaboration (B.B. Back et al.). Sep 2003. 6pp. 
%Published in Phys.Rev.C70:011901,2004 
%e-Print Archive: nucl-ex/0309013 

%CHARGED ANTIPARTICLE TO PARTICLE RATIOS NEAR MIDRAPIDITY IN P + P COLLISIONS AT S(NN)**(1/2) = 200-GEV.
%By PHOBOS Collaboration (B.B. Back et al.). Sep 2004. 3pp. 
%Published in Phys.Rev.C71:021901,2005 
%e-Print Archive: nucl-ex/0409003 

%33
\bibitem{BRAHMS}
I.G.~Bearden {\it et al.}, [BRAHMS Collaboration], \PLB{\bf 607}, 42 (2005).
%FORWARD AND MIDRAPIDITY LIKE-PARTICLE RATIOS FROM P + P COLLISIONS AT S**(1/2) = 200-GEV.
%By BRAHMS Collaboration (I.G. Bearden et al.). Sep 2004. 13pp. 
%Published in Phys.Lett.B607:42-50,2005 
%e-Print Archive: nucl-ex/0409002 


%34
\bibitem{ppg028} S.S. Adler {\it et al.}, [PHENIX Collaboration],
			\PRL{\bf 91}, 072303 (2003).

%35
\bibitem{starpiddau} 
J.~Adams {\it et al.}, [STAR Collaboration], \PLB{\bf 616}, 8 (2005).
%J. Adams {\it et al.} [STAR Collaboration], nucl-ex/0309012 (2003).


%36
\bibitem{boris} B.Z. Kopeliovich, J. Nemchik, A. Schafer and
A.V. Tarasov, \PRL{\bf 88}, 232303 (2002).

\end{thebibliography}
\end{document}